\documentclass[aps,prb,twocolumn,superscriptaddress,longbibliography]{revtex4-2}
\usepackage{amsmath,amssymb}
\usepackage[pdftex]{hyperref,graphicx}
\hypersetup{colorlinks = true, urlcolor = blue, linkcolor = blue, citecolor = blue}
\usepackage{physics}
\usepackage{xcolor}
\usepackage{bm}
\usepackage{float}
\usepackage{dsfont}

\begin{document}
\title{Local and nonlocal stochastic control of quantum chaos: Measurement- and~control-induced criticality}
\author{Haining Pan}
\affiliation{Department of Physics and Astronomy, Center for Materials Theory, Rutgers University, Piscataway, NJ 08854 USA}

\author{Sriram Ganeshan} 
\affiliation{Department of Physics, City College, City University of New York, New York, NY 10031, USA}
\affiliation{CUNY Graduate Center, New York, NY 10031}

\author{Thomas Iadecola}
\affiliation{Department of Physics and Astronomy, Iowa State University, Ames, IA 50011, USA}
\affiliation{Ames National Laboratory, Ames, IA 50011, USA}

\author{Justin H. Wilson}
\affiliation{Department of Physics and Astronomy, Louisiana State University, Baton Rouge, LA 70803, USA}
\affiliation{Center for Computation and Technology, Louisiana State University, Baton Rouge, LA 70803, USA}

\author{J. H. Pixley}
\affiliation{Department of Physics and Astronomy, Center for Materials Theory, Rutgers University, Piscataway, NJ 08854 USA}
\affiliation{Center for Computational Quantum Physics, Flatiron Institute, New York, NY 10010, USA}
\date{\today}

\begin{abstract}
    We theoretically study the topology of the phase diagram of a family of quantum models inspired by the classical Bernoulli map under stochastic control. 
    The quantum models inherit a control-induced phase transition from the classical model and also manifest an entanglement phase transition intrinsic to the quantum setting. 
    This measurement-induced phase transition has been shown in various settings to either coincide or split off from the control transition, but a systematic understanding of the necessary and sufficient conditions for the two transitions to coincide in this case has so far been lacking. 
    In this work, we generalize the control map to allow for either local or global control action. 
    While this does not affect the classical aspects of the control transition that is described by a random walk, it significantly influences the quantum dynamics, leading to the universality class of the measurement-induced transition being dependent on the locality of the control operation.
    In the presence of a global control map, the two transitions coincide and the control-induced phase transition dominates the measurement-induced phase transition.
    Contrarily, the two transitions split in the presence of the local control map or additional projective measurements and generically take on distinct universality classes. 
    For local control, the measurement-induced phase transition recovers the Haar logarithmic conformal field theory universality class found in feedback-free models.
    However, for global control, a novel universality class with correlation length exponent $\nu \approx 0.7$ emerges from the interplay of control and projective measurements.
    This work provides a more refined understanding of the relationship between the control- and measurement-induced phase transitions.
\end{abstract}
\maketitle

\section{Introduction}
The dynamics of a quantum many-body system under local unitary evolution leads to a volume-law entangled steady state where the entanglement between a subsystem and its complement scales with the subsystem volume~\cite{nahum2017quantum}.
Adding local projective measurements tends to remove entanglement from the system and can drive a transition to an area-law entangled steady state, where the entanglement scales with the subsystem boundary. This measurement-induced phase transition (MIPT)~\cite{skinner2019measurementinduced,li2018quantum,li2019measurementdriven,chan2019unitaryprojective,potter2022entanglement,fisher2023random} between volume- and area-law entangled steady states can be probed by higher moments of observables~\cite{li2019measurementdriven} or by entanglement measures, such as the tripartite mutual information~\cite{zabalo2020critical}.
However, these quantities are not easily accessible in experiments because they are not linear in the density matrix and therefore only take nontrivial average values upon resolving individual quantum trajectories corresponding to different measurement histories.
This overhead, exponential in the number of intermediate measurements, results in the so-called ``postselection problem'' to observe the MIPT.

Recent experimental efforts to directly observe the MIPT have utilized Clifford gates with a classical decoder~\cite{noel2022measurementinduced} as well as resolving the measurement histories by brute force, which is not scalable~\cite{koh2023measurementinduced,hoke2023measurementinduced}. 
Theoretical proposals to observe the MIPT aim to circumvent the postselection problem using different approaches to ``linearize" the calculation, such as the cross-entropy benchmark~\cite{li2023cross} and quantum estimators like shadow tomography~\cite{garratt2023measurements,garratt2023probing,dehghani2023neuralnetwork,ippoliti2024learnability}. However, these proposals require access to a reference dynamics run on a classical simulator, which then limits the observability of the MIPT to settings like Clifford circuits where classical simulations are scalable.

Motivated in part by this challenge, recent works have considered introducing feedback operations conditioned on the measurement outcomes~\cite{iadecola2023measurement,odea2024entanglement,buchhold2022revealing,ravindranath2023entanglement,piroli2023triviality,sierant2023controlling}.
The quantum channel comprised of measurements and feedback can be viewed as a ``control map" that attempts to steer the system's dynamics onto a preselected steady state.
This can lead to a control-induced phase transition (CIPT, also known as an absorbing-state phase transition) above which the system reaches the target state regardless of the initial condition.
While these CIPTs manifest in local order parameters and correlation functions, and therefore are experimentally observable, they generally occur separately from the MIPT even when measurements are only applied in concert with feedback~\cite{odea2024entanglement,ravindranath2023entanglement}.
It is therefore desirable to understand whether and when the MIPT and CIPT can coincide, so that the latter can serve as an experimentally accessible indicator for the former.

In this paper, we study this question from a vantage point grounded in the theory of classical dynamical systems, where CIPTs arise in the so-called probabilistic control of chaos~\cite{antoniou1996probabilistic,antoniou1997probabilistic,antoniou1998absolute}.
In this setup, a (classical) chaotic map is stochastically interleaved with a control map that attempts to stabilize an unstable trajectory of the chaotic dynamics.
At each time step, the control is applied with probability $p_{\rm ctrl}$, and otherwise the chaotic dynamics is applied; the CIPT occurs above a critical value $p^c_{\rm ctrl}$.
Inspired by this protocol, Ref.~\cite{iadecola2023measurement} considered a quantum circuit model inspired by the classically chaotic Bernoulli map~\cite{Renyi1957representations}, where the simplest classical example of a CIPT occurs~\cite{antoniou1996probabilistic,antoniou1998absolute}.
There, it was shown that the classical CIPT persists in the quantum model and coincides with an MIPT.
However, Ref.~\cite{lemaire2024separate} developed a Clifford version of this model and found that the MIPT and CIPT separate, with the former preceding the latter.
Aside from the restriction to Clifford circuits, one key difference between Refs.~\cite{lemaire2024separate} and~\cite{iadecola2023measurement} is that the former used a local control map while the latter used a long-range one.
While this change was originally implemented for technical reasons, it may have profound implications for the relationship between measurement- and control-induced criticality.
For example, Ref.~\cite{sierant2023controlling} demonstrated using Clifford-circuit simulations that a long-ranged control map tends to align the MIPT with the CIPT, while a short-ranged control map results in two distinct phase transitions.

Motivated by this observation, in this paper, we undertake a systematic study of the impact of the control map's structure on the interplay of the MIPT and CIPT in the Bernoulli circuit model of Ref.~\cite{iadecola2023measurement}.
We find that the structure of the control map does not influence the CIPT,
but strongly influences the MIPT.
We confirm that, while a globally acting control map can push the MIPT and CIPT together, a locally acting one can pull them apart, forcing the MIPT into a different universality class while leaving the CIPT unchanged.
We also find that incorporating additional projective measurements into the dynamics can allow the two transitions to be tuned continuously.
Our numerical results show that the MIPT for the local control map manifests a Haar logarithmic conformal field theory (log-CFT) universality class~\cite{zabalo2022operator}. Whereas, in the limit of the zeroth Renyi entropy (that is equivalent to taking the onsite Hilbert space $d$ to infinity~\cite{skinner2019measurementinduced,jian2020measurementinduced,agrawal2022entanglement}) our numerical results are consistent with the recent findings from an effective statistical mechanics model in the $d\rightarrow \infty$ for much larger system sizes~\cite{allocca2024statistical}.
The latter effective model, which applies in the limit of infinite onsite Hilbert space dimension, finds that the CIPT and MIPT always coincide even for a local control map. This feature manifests itself in our finite-$d$ model of qubits in the behavior of the zeroth R\'enyi entropy.

The remainder of the paper is organized as follows.
In Sec.~\ref{sec:model}, we introduce the Bernoulli map and its quantum analog, outline the six different types of local and nonlocal control operations we will consider, and define the metrics used to detect the CIPT and MIPT.
In Sec.~\ref{sec:coincide}, we focus on the scenario where the two transitions coincide.
In Sec.~\ref{sec:split}, we split the transitions, either by modifying the structure of the control map (Sec.~\ref{sec:local_adder}) or by adding projective measurements without additional control (Sec.~\ref{sec:projection}).
We also provide evidence in Sec.~\ref{sec:S0} that, in the infinite-$d$ limit with local control, the MIPT and CIPT coincide again.
In Sec.~\ref{sec:phase_diagram}, we discuss the distinctive topology of the phase diagram after incorporating these two modifications.
In Sec.~\ref{sec:nature}, we discuss the crucial factors that determine the topology of the phase diagram.
We conclude in Sec.~\ref{sec:conclusion}.
In Appendix~\ref{app:datacollapse} we present the details of the data collapse, and in Appendix~\ref{app:kraus} we summarize the Kraus-operator representations of the various types of control we consider.
\section{Models and Approach}\label{sec:model}

\begin{figure*}[ht]
    \centering
    \includegraphics[width=6.8in]{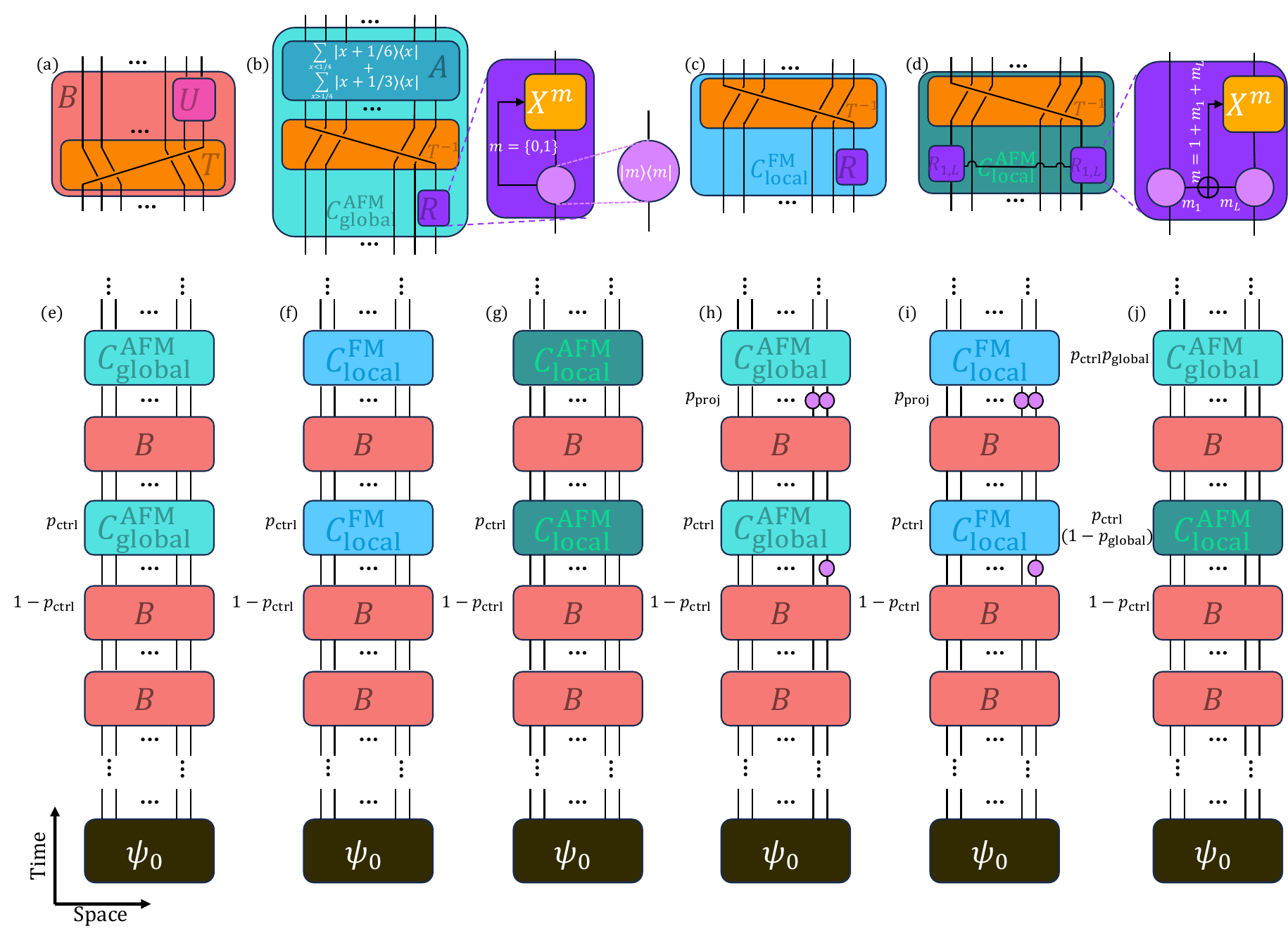}
    \caption{
    (a) The Bernoulli map $B$ is composed of a left-shift operator $T$ followed by a Haar random unitary operator $U$ acting on the last two qubits [see Eq.~\eqref{eq:B}].
    (b) The global control map for the period-two AFM orbit, $C_{\text{glolbal}}^{\text{AFM}}$, is composed of a reset $R$ on the last qubit followed by a right-shift operator $T^{-1}$ and an adder $A$ (see Eq.~\eqref{eq:C_global_AFM}). 
    (c) The local control map for the FM fixed point, $C_{\text{local}}^{\text{FM}}$, removes the adder $A$ (see Eq.~\eqref{eq:C_local_FM}). 
    (d) The local control map for the AFM orbit $C_{\text{local}}^{\text{AFM}}$ uses a two-qubit projector in the reset [see Eq.~\eqref{eq:C_local_AFM}]. 
    Bottom panels show examples of stochastic quantum circuits for (e) the global control map with AFM fixed points, (f) the local control map with FM fixed points, (g) the local control map with AFM fixed points, (h) the global control map with AFM fixed points and additional projective measurements, (i) the local control map with FM fixed points and additional projective measurements, and (j) the interpolation between the local and global control map with AFM fixed points. Kraus operators for the various control maps are given in Table~\ref{table:kraus} of Appendix~\ref{app:kraus}.
    }
    \label{fig:schematic}
\end{figure*}

\subsection{Probabilistic control of a classical Bernoulli map}

We start with a model of quantum dynamics~\cite{iadecola2023measurement} where one stochastically applies one of two competing operations: a quantum circuit analog of the Bernoulli map $B$ (with probability $1-p_{\text{ctrl}}$) and the control map $C$ (with probability $p_{\text{ctrl}}$). 

This model originates from the field of classical dynamical systems, where the chaotic dynamics of the classical Bernoulli map were stochastically controlled by the introduction of a control map \cite{antoniou1996probabilistic}. 
The Bernoulli map $B$ is defined by the operation 
\begin{equation}
    B:x \mapsto 2x~\text{mod}~1,
\end{equation}
where $x\in[0,1)$.
Any rational number $x \in [0,1)$ undergoes a finite-length periodic orbit under this map.
However, for irrational $x$, the dynamics are chaotic; thus, since any rational $x$ is arbitrarily close to an irrational number, these periodic orbits are unstable.
The control map $C$ aims to stabilize these unstable orbits.

Suppose we want to target an orbit consisting of the points $x_F=\left\{x_f^{(1)},x_f^{(2)},\dots\right\}$.
Then the control map acts on $x\in[0,1)$ as
\begin{equation}
    C:x \mapsto(1-a) x_f +a x,
\end{equation}
where $x_f\in x_F$ is the point on the orbit that is closest to $x$ and $a$ sets the strength of the control.
Iterating $C$ on $x$ leads to the fixed point $x_f$ for any $|a|<1$---in other words, the control map has fixed points corresponding to each point on the target orbit.
The control map tends to counteract the chaotic dynamics generated by $B$ and, if applied sufficiently frequently, leads to a controlled phase where the target orbit is reached from any initial condition.
We set $a=\frac{1}{2}$ to have a CIPT at $p_{\text{ctrl}}^c=0.5$~\cite{antoniou1997probabilistic,antoniou1998absolute}, below which the system is in the chaotic phase, and above which the system is in the controlled phase.
As discussed in Ref.~\cite{iadecola2023measurement}, this phase transition is described by an unbiased random walk with dynamical exponent $z=2$ and correlation length critical exponent $\nu=1$.

\subsection{Quantum analog of the Bernoulli map}
In the quantum model, we first digitize any real number $x \in [0,1)$ into a binary representation by truncating it to $L$ bits, and encoding the resulting bit string into a computational basis (CB) of $L$ qubits:
\begin{equation}
    \left( x \right)_{10}=\left( {0.b_1b_2\dots b_L} \right)_2=\ket{b_1b_2\dots b_L}.
\end{equation}

To simulate the Bernoulli map $B$ in this quantum system,  we apply a cyclic leftward shift operator $T$ to implement multiplication by 2, i.e.,
\begin{equation}
    T\ket{b_1b_2\dots b_L}= \ket{b_2\dots b_Lb_1}.
\end{equation} 
Since $T$ shifts the leftmost qubit into the rightmost position, it generates an orbit of length at most $L$ for any $x$.
To recover the chaotic phase, we apply a unitary scrambling operation $U$ to qubits $L-1$ and $L$ after each application of $T$. 
This scrambling operation is what produces nontrivial quantum dynamics in the model; in this work, we take it to be a random unitary drawn from the Haar measure on U(4).
The full Bernoulli circuit is then implemented by the unitary operator [see Fig.~\ref{fig:schematic}(a)]
\begin{equation}\label{eq:B}
    B=U T.
\end{equation}

\subsection{Control map}
The control map $C$ in the quantum model is implemented in two steps: The first step is to halve $x$, which can be realized by the right-shift operator $T^{-1}$.
However, since the rightmost qubit will be shifted to the leftmost position after applying $T^{-1}$, we need to ensure that the leftmost qubit will always be in the state $\ket 0$. 
This is achieved by resetting the last qubit to zero before applying $T^{-1}$. 
This reset $R$ is implemented by first measuring the qubit
\begin{equation}\label{eq:M}
    \mathcal{M}_{P_i^{m}}\ket{\psi}=\frac{P_i^m\ket{\psi}}{\norm*{P_i^m\ket{\psi}}}, 
\end{equation}
where 
\begin{equation} \label{eq:Pim}
    P_i^{m}=\ketbra{m}_i,
\end{equation}
(we always normalize the state after projective measurements) with $m=0,1$ and then applying a Pauli $X$ gate if its measurement outcome $m=1$ [see Fig.~\ref{fig:schematic}(c)]:
\begin{equation}
    R_L=(X_L)^{m}P_L^m,
\end{equation}
where $(X_L)^{m}$ is the $m$th power of the Pauli matrix $X$ acting on the $L$-th qubit.

The second step of the control is to add a fixed value determined by the fixed points $x_F$.
Namely, we have an adder operator 
\begin{equation}
    A=\sum_{x\in\Delta_{x_f}}\sum_{x_f\in x_F}\ketbra{x+x_f/2}{x}.
\end{equation}
Here, $\Delta_{x_f}$ is a neighborhood of the fixed point $x_f$ consisting of all points that will be attracted to $x_f$ under the control, chosen such that $\cup_{x_f\in x_F}\Delta_{x_f} = \mathbb{R}\cap [0,1)$.

In this paper, we consider three types of adders that steer the dynamics to two sets of different fixed points: (i) the global adder with $x_F=\left\{ 1/3,2/3 \right\}$ as shown in Fig.~\ref{fig:schematic}(b), (ii) the local adder with $x_F=\{0\}$ as shown in Fig.~\ref{fig:schematic}(c), and (iii) the local adder with $x_F=\left\{ 1/3,2/3 \right\}$ as shown in Fig.~\ref{fig:schematic}(d).
The global adder has a support scaling with the system size $L$, while the local adders have a fixed support that is independent of the system size.
Intuitively, in the limit $L\to\infty$, the global adder adds a number whose binary expansion contains infinitely repeating bits, while the local adder adds a number with a binary expansion of finite length. We now describe the three adder circuits in detail below.

\subsubsection{Global control with $x_F=\left\{ 1/3,2/3 \right\}$}
For the orbit $x_F=\left\{ 1/3,2/3 \right\}$~\cite{iadecola2023measurement}, the adder operator $A$ is
\begin{equation}
    A=\sum_{x<1/4}\ketbra{x+1/6}{x}+\sum_{x\ge1/4}\ketbra{x+1/3}{x},
\end{equation}
which means that we add $1/6$ for any $x<1/4$ and $1/3$ for any $x\ge1/4$. 
(Note that $x$ is only defined within $[0,1/2)$ after halving $T^{-1}$.)
In the binary representation, these two fixed points are $\left( \frac{1}{6} \right)_{10}=\left( 0.0\overline{01} \right)_2$ and $\left( \frac{1}{3} \right)_{10}=\left( 0.\overline{01} \right)_2$, so it is a global adder as the unstable fixed points require an infinite bit string representation that spans the full system size when truncated.

Combining this adder operator with $T^{-1}$ yields a global control map, which maps the fixed points $x_F=\left\{ 1/3,2/3 \right\}$ to themselves. 
In the CB, these two fixed points take the form of the two antiferromagnetic (AFM) N\'eel states: $\ket{01}^{\otimes L/2}$ and $\ket{10}^{\otimes L/2}$ (assuming $L\in2\mathbb{Z}^+$). 
We denote this control map as (up to the wave function normalization)
\begin{equation}\label{eq:C_global_AFM}
    C_{\text{global}}^{\text{AFM}}=A T^{-1}R_L=A T^{-1}(X_L)^mP_L^m
\end{equation} 
as shown in Fig.~\ref{fig:schematic}(b). 
An example stochastic quantum circuit in which this control map competes with the Bernoulli circuit is shown in Fig.~\ref{fig:schematic}(e).

\subsubsection{Local control with $x_F=\left\{ 0 \right\}$}

The two other types of control maps we consider use a local adder.
The simplest version is the identity operator, which effectively adds 0 (i.e., $\ket{x}\mapsto \ket{x\oplus 0}=\ket{x}$), and leads to a single fixed point $x_F=\left\{ 0 \right\}$, which is a ferromagnetic (FM) state ($\ket{0}^{\otimes L}$) in the CB representation.
This essentially removes the adder from the control map (as its support is trivially zero, independent of the system size), i.e., 
\begin{equation}\label{eq:C_local_FM}
    C_{\text{local}}^{\text{FM}}=\mathds{1}T^{-1}R_L=T^{-1}(X_L)^mP_L^m,
\end{equation}
as shown in Fig.~\ref{fig:schematic}(c).
An example stochastic quantum circuit pitting this local adder against the Bernoulli circuit is shown in Fig.~\ref{fig:schematic}(f).
In Sec.~\ref{sec:split}, we will show that the MIPT can be separated from the CIPT by replacing the global control map with this local control map.

\subsubsection{Local control with $x_F=\left\{ 1/3,2/3 \right\}$}

Finally, we propose a local adder that can control onto the same orbit as the global adder, $x_F=\left\{ 1/3,2/3 \right\}$, as shown in Fig.~\ref{fig:schematic}(d).
Here, we replace the reset $R$ in Eq.~\eqref{eq:C_global_AFM} with a different conditional operation acting on two qubits instead of one.
Namely, for a chain with even $L$, we perform Born-rule projections on both the first and last qubit. 
The new conditional feedback operation flips the last qubit if the measurement outcomes for the first and last qubit are the same, i.e., 
\begin{equation}
    R_{1,L}=(X_L)^{m_1+m_L+1} P_{1}^{m_1}P_{L}^{m_L}
\end{equation}
up to a normalization factor after the projection, where $m_i=\left\{ 0,1 \right\}$ is the measurement outcome for qubit $i$.
It is easy to verify that this conditional operation removes the need for a global adder, controlling onto the same fixed points as the global adder.
We denote the full control map by (up to the wavefunciton normalizatoin)
\begin{equation}\label{eq:C_local_AFM}
    C_{\text{local}}^{\text{AFM}} =\mathds{1}T^{-1}R_{1,L}=  T^{-1} (X_L)^{{m_1+m_L+1}} P_{1}^{m_1}P_{L}^{m_L}.
\end{equation}
The corresponding stochastic quantum circuit is shown in Fig.~\ref{fig:schematic}(j).
Note that, unlike $C_{\text{global}}^{\text{AFM}}$, this control operation does not have fixed points at $1/3$ and $2/3$, but rather  cycles through the orbit $1/3\leftrightarrow 2/3$ through the two maps $x\mapsto x/2+1/2$ for $x< 1/2$ and $x\mapsto x/2$ for $x\ge 1/2$.
In Sec.~\ref{sec:local_AFM}, we will show that this local control map also splits the MIPT from the CIPT.
Finally, we note that this control protocol can be generalized to target other orbits consisting of CB states with repeating patterns of bits.

\subsection{Adding projective measurements}
\label{sec:model-proj}
Another way to split the MIPT and CIPT is to introduce projective measurements without feedback, i.e., Eq.~\eqref{eq:M} and Eq.~\eqref{eq:Pim}.
Here, projective measurements are stochastically applied to qubits $L-1$ and $L$ with probability $p_{\rm proj}$ after the unitary $U$, as represented by the purple dots in Fig.~\ref{fig:schematic}(h) for global control map, and Fig.~\ref{fig:schematic}(i) for local control map.
We stress that these feedback-free projective measurements are part of the chaotic map and that the tuning parameters $p_{\rm ctrl}$ and $p_{\rm proj}$ are independent.

\subsection{Interpolation between the global and local adder}
Since the global control $C_{\text{global}}^{\text{AFM}}$ and the local control $C_{\text{global}}^{\text{AFM}}$ both target the same orbit $x_F=\left\{ 1/3,2/3 \right\}$, we can randomly choose between them at each control step as shown in Fig.~\ref{fig:schematic}(j), where the global control map is applied with probability $p_{\text{global}}$, and the local control map with probability $1-p_{\text{global}}$.
This provides a smooth interpolation between the global and local adders to study the effect of the degree of locality of the control map.

To summarize, we have six different types of stochastic quantum circuits [Fig.~\ref{fig:schematic}(e)--(j)], distinguished by the different types of control map summarized above. We define these control operations using the language of Kraus operators in Table~\ref{table:kraus} of Appendix~\ref{app:kraus}.

\subsection{Metrics to probe the CIPT and MIPT}

\subsubsection{Order parameters for CIPT}
To probe the CIPT, we use a macroscopic observable as an order parameter, defined such that it approaches its maximal value of 1 in the controlled phase and zero in the chaotic phase as system size $L\to\infty$. 
The critical point is then determined by the finite-size crossing of the order parameter as it interpolates between these two behaviors.

For the period-2 orbit $x_F=\{1/3,2/3\}$, we adopt the classical N\'eel order parameter 
\begin{equation}\label{eq:O_AFM}
    O_{\text{AFM}}=-\frac{1}{L}\sum_{i=1}^L Z_i Z_{i+1},
\end{equation}
which detects the AFM order manifested by the CB representation of the states on the orbit.
Here, we impose periodic boundary conditions (i.e., $Z_{L+1}\equiv Z_{1}$) and denote by $Z_i$ the Pauli $z$ matrix ($\sigma_z$) acting on the $i$-th qubit (i.e., $Z_i\ket{b_i}=(-1)^{b_i}\ket{b_i}$ for $b_i\in\{0,1\}$).

For the FM fixed point $x_F=\{0\}$, we use the order parameter 
\begin{equation}\label{eq:O_FM}
    O_{\text{FM}}=\frac{1}{L}\sum_{i=1}^L Z_{i},
\end{equation}
 which is maximized by the CB state $\ket{0}^{\otimes L}$.

We compute the order parameter $O_{\text{AFM/FM}}$ for each quantum trajectory to obtain the quantum expectation value $\expval{O_{\text{AFM/FM}}}$. Each quantum trajectory is described by a pure state $\rho_{\vec m}=|\psi_{\vec{m}}\rangle\langle \psi_{\vec{m}}|$, where $\vec m$ denotes the full record of all measurement outcomes obtained during the course of the evolution (including both conditional feedback operations and any additional feedback-free projective measurements).
We also take each quantum trajectory to have its own realization of the stochastic circuit.
We then average over all quantum trajectories $\vec m$ to obtain $\overline{\expval{O_{\text{AFM/FM}}}}=\tr(O_{\text{AFM/FM}}\bar{\rho})$, where $\bar{\rho}=\sum_{\vec{m}}\rho_{\vec{m}}$ is the average density matrix.
Note that these averages commute because the observables are linear in the density matrix.

\subsubsection{Tripartite mutual information for MIPT}
\begin{figure*}[!ht]
    \centering
    \includegraphics[width=6.8in]{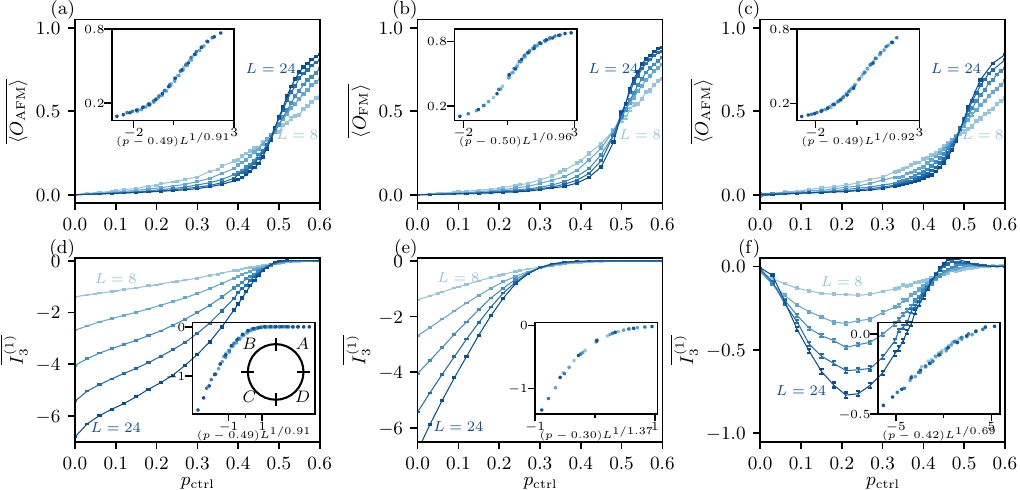}
    \caption{
    Top panels: The order parameter [see Eqs.~\eqref{eq:O_AFM} and~\eqref{eq:O_FM}] as a function of $p_{\text{ctrl}}$ for (a) a global adder with $x_F=\left\{ 1/3,2/3 \right\}$; (b) a local adder with $x_F=\{0\}$; (c) a global adder with $x_F=\left\{ 1/3,2/3 \right\}$ at the projective measurement rate of $p_{\text{proj}}=0.3$. 
    The insets show the data collapse near the critical point at (a) $p_{\text{ctrl}}=0.488(1)$, (b) $p_{\text{ctrl}}=0.497(1)$, and (c) $p_{\text{ctrl}}=0.485(2)$ for $L\ge12$.
    Bottom panels: The tripartite mutual information $\overline{I_3^{(1)}}$ [see Eq.~\eqref{eq:I3}] as a function of $p_{\text{ctrl}}$ for (d) a global adder with $x_F=\left\{ 1/3,2/3 \right\}$; (e) a local adder with  $x_F=\{0\}$; (f) a global adder with $x_F=\left\{ 1/3,2/3 \right\}$ at the projective measurement rate of $p_{\text{proj}}=0.3$. The insets show the data collapse near the critical point at (d) $p_{\text{ctrl}}=0.485(3)$; (e) $p_{\text{ctrl}}=0.297(2)$; (f) $p_{\text{ctrl}}=0.416(2)$.
    The geometry of the tripartite mutual information is shown in the inset of (d).
    The system size is a multiple of 4 from $L=8$ to 24 and the ensemble size is 2000.
    }
    \label{fig:linecuts}
\end{figure*}

To probe the MIPT, we use the tripartite mutual information~\cite{kitaev2006topological,zabalo2020critical}.
We divide the system into four subregions with equal lengths: $A=[1, L/4]$, $B=[L/4+1, L/2]$, $C=[L/2+1,3L/4]$, and $D=[3L/4+1, L]$ [see schematic in the inset of Fig.~\ref{fig:linecuts}(d)].
We ensure that the system size $L$ is a multiple of 4 to avoid boundary effects.
The tripartite mutual information $I_3^{(n)}$ is then defined as 
\begin{equation}\label{eq:I3}
    I_3^{(n)}=S_A^{(n)}+S_B^{(n)}+S_C^{(n)}-S_{A\cup B}^{(n)}-S_{B\cup C}^{(n)}-S_{A\cup C}^{(n)}+S_{A\cup B\cup C}^{(n)},
\end{equation}
where $S_{i}^{(n)}$ is $n$-th R\'enyi entropy of the reduced density matrix of the subregion $i$, i.e.,
\begin{equation}
    S_{i}^{(n)}=\frac{1}{1-n}\log\tr(\rho_i^n).
\end{equation}
In the limit $n\to 1$ we recover the von Neumann entanglement entropy,
\begin{equation}
    S_{i}^{(1)}=-\tr(\rho_i\log\rho_i),
\end{equation}
while for $n=0$ we recover the Hartley entropy,
\begin{equation}
    S_{i}^{(0)}=\log\tr(\rho_i^0).
\end{equation}
Importantly for qubits, the zeroth Renyi entropy is equivalent to taking the onsite Hilbert space $d\rightarrow \infty$~\cite{skinner2019measurementinduced,jian2020measurementinduced,agrawal2022entanglement} and therefore the Hartley entropy $S_{i}^{(0)}$ allows us to make contact with recent results on a statistical mechanics model in this limit~\cite{allocca2024statistical}.

We use $I_3^{(1)}$ to detect the MIPT, and $I_3^{(0)}$ to detect the percolation transition in Sec.~\ref{sec:S0}.
We compute them for each quantum trajectory and then average over trajectories (note that these averages do not commute because $I_3^{(n)}$ is not linear in the density matrix.
$I_3^{(n)}<0$~\cite{zabalo2020critical} should scale as the system size $L$ in the volume-law phase, and saturate to a constant in the area-law phase.
The MIPT critical point is thus indicated by a finite-size crossing in $I_3^{(n)}$.

In the following, we will first revisit the model with fixed points $x_F=\left\{ 1/3,2/3 \right\}$, where the CIPT and MIPT coincide, in Sec.~\ref{sec:coincide}.
In Sec.~\ref{sec:split} we will then separate the CIPT and MIPT by either modifying the locality of the adder $A$ or introducing feedback-free projective measurements.

\section{Global control map with AFM fixed points}\label{sec:coincide}

In this section, we demonstrate the scenario where the CIPT and MIPT overlap using a global control map with fixed points $x_F=\left\{ 1/3,2/3 \right\}$ as shown in Fig.~\ref{fig:schematic}(e).
We evolve the circuit of Fig.~\ref{fig:schematic}(e) for $2L^2$ steps to ensure that it enters the steady state.
Figure~\ref{fig:linecuts}(a) shows the trajectory-averaged order parameter $\overline{\expval{O_{\text{AFM}}}}$.
We find that the CIPT happens at $p_{\text{ctrl}}\approx 0.5$, in agreement with the classical result and with Ref.~\cite{iadecola2023measurement} in the quantum limit.

To detect the MIPT, we plot in Fig.~\ref{fig:linecuts}(d) the trajectory-averaged tripartite mutual information $\overline{I_3^{(1)}}$ as a function of $p_{\text{ctrl}}$ for different system sizes and find that the critical control rate is also at $p_{\text{ctrl}}\approx0.5$.
This demonstrates the idea of using the CIPT (witnessed by an observable linear in the density matrix) to herald the MIPT.

In addition, we perform data collapse (see Appendix~\ref{app:datacollapse} for more details) and find that the correlation length critical exponent of the MIPT shows a value of $\nu=0.90(5)$ [as shown in the inset of Fig.~\ref{fig:linecuts}(d)] as opposed to the Haar log-CFT universality class with $\nu\approx 1.3$~\cite{zabalo2020critical}, which is in good agreement with Ref.~\cite{iadecola2023measurement} (as expected) but the present results extend the study now to system sizes of $L=24$.
This critical exponent is consistent with the CIPT [as shown in the inset of Fig.~\ref{fig:linecuts}(a)], indicating that the random-walk criticality of the CIPT dominates over that of the MIPT~\cite{iadecola2023measurement,allocca2024statistical}.
However, this random-walk criticality with $\nu\approx 1$ is not robust to perturbations that split the two transitions: when the CIPT is separated from the MIPT, the critical exponent of the MIPT returns to the Haar log-CFT universality class with $\nu\approx 1.3$.
In the following sections, we will discuss the scenarios where the two transitions are split.

\section{Splitting the transitions: Local control and projective measurements}\label{sec:split}
In this section, we demonstrate two ways to split CIPT and MIPT: replacing the global adder with a local adder, and introducing feedback-free projective measurements. 
We also investigate signatures of the CIPT and MIPT in the tripartite mutual information based on the zeroth R\'enyi entropy, which is sensitive to the behavior in the limit of infinite onsite Hilbert space dimension.
We then present a thorough exploration of the full phase diagrams for global and local control with and without projective measurements.

\subsection{Local adder with $x_F=\left\{ 0 \right\}$}\label{sec:local_adder}

Replacing the global adder with the local adder separates the MIPT from the CIPT.
Figure~\ref{fig:linecuts}(b) shows that the CIPT remains at $p_{\text{ctrl}}=0.498(1)$ with the same random-walk universality class showing a critical exponent of $\nu=0.96(3)$. 
However, for the MIPT, we plot the tripartite mutual information as shown in Fig.~\ref{fig:linecuts}(e) and find that the critical control rate now decreases to a lower value of $p_{\text{ctrl}}^c=0.297(2)$ with a larger critical exponent of $\nu=1.36(4)$ [as shown in the inset of Fig.~\ref{fig:linecuts}(e)], which recovers the Haar log-CFT universality class~\cite{zabalo2020critical}.
This implies that the nature of the adder (global versus local) plays a crucial role in determining the splitting of the CIPT and MIPT, affecting critical properties of the MIPT. 
Before we delve into the fundamental reasons for this change, we propose another way to split the two transitions, i.e., by introducing feedback-free projective measurements.

\subsection{Feedback-free projective measurements}\label{sec:projection}

The second way to split CIPT and MIPT is to introduce feedback-free projective measurements with probability $p_{\text{proj}}$ into the chaotic map as described in Sec.~\ref{sec:model-proj}.
In the limit $p_{\rm ctrl}\to 0$, this should recover the usual Haar log-CFT transition~\cite{skinner2019measurementinduced,li2019measurementdriven} as a function of $p_{\text{proj}}$ with the only difference being that the unitary gates are applied in a staircase fashion rather than a bricklayer fashion. 
In the limit of zero measurement rate, $p_{\text{proj}}=0$, the CIPT and MIPT coincide at $p_{\text{ctrl}}\approx 0.5$, as shown in Fig.~\ref{fig:linecuts}(a). 
In the intermediate regime between $p_{\text{proj}}=0$ and $p_{\text{ctrl}}=0$, we expect the critical point of the MIPT to shift away from the CIPT. 

We present numerical results at a finite measurement rate $p_{\text{proj}}=0.3$ in Fig.~\ref{fig:linecuts}(c,f).
In Fig.~\ref{fig:linecuts}(c), we find that the CIPT remains at $p_{\text{ctrl}}=0.486(1)$ with the same random-walk universality class displaying a critical exponent $\nu=0.92(3)$.
This indicates that the finite measurement rate does not affect the CIPT at all.
In Fig.~\ref{fig:linecuts}(f), we plot the tripartite mutual information as a function of $p_{\text{ctrl}}$ at $p_{\text{proj}}=0.3$, and find that there are now two critical points. 
As $p_{\text{ctrl}}$ increases from 0, the system undergoes a transition from area-law to volume-law scaling at $p_{\text{ctrl}}\approx 0.05$, and then returns to the area-law phase at $p_{\text{ctrl}}^c=0.416(2)$, which is smaller than the previous critical control rate of $p_{\text{ctrl}}^c=0.485(2)$.
The initial area-law scaling arises because the projective measurements at $p_{\rm proj}=0.3$ are sufficiently strong that the system is above the MIPT into the area-law phase already.
The MIPT also manifests a different critical exponent around $\nu=0.70(3)$. 
We will discuss this new universality later in Sec.~\ref{sec:phase_diagram}.

\subsection{Zeroth R\'enyi entropy}\label{sec:S0}

The zeroth R\'enyi entropy (also known as the Hartley entropy) tracks physics not captured with previous measures; in monitored random circuits without feedback, it is controlled by a different phase transition entirely: percolation ~\cite{nahum2017quantum,skinner2019measurementinduced}.
For a subset of qubits in these models, this is precisely related to the bond dimension and is thereby mapped onto a percolation transition for the 1+1D circuit.
It also happens to be the only MIPT obtained in the limit of infinite onsite Hilbert space with Haar gates~\cite{jian2020measurementinduced}, whereas in the stabilizer limit it depends on how the large-$d$ limit is taken~\cite{li2024statistical}.
Similar to $I_{3}^{(1)}$, a finite-size crossing of $I_3^{(0)}$ indicates the percolation transition in bond dimension in monitored random circuits.
This transition occurs generically after the area-law phase but indicates that entanglement is compactly localized, we, therefore, call it ``compact area-law''
Due to the intimate connection to percolation, this measure can be benchmarked with much larger numerics which simulate infinite on-site dimension~\cite{allocca2024statistical}.

\begin{figure}[ht]
    \centering
    \includegraphics[width=3.4in]{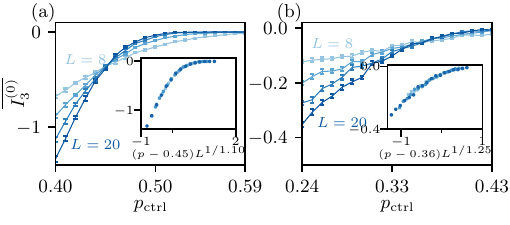}
    \caption{
        The tripartite mutual information $\overline{I_3^{(0)}}$ [see Eq.~\eqref{eq:I3}] with zeroth R\'enyi entropy as a function of $p_{\text{ctrl}}$ for (a) $p_{\text{proj}}=0$, and (b) $p_{\text{proj}}=0.36$. The insets show the data collapse near the critical point (a) $p_{\text{ctrl}}=0.453(2)$ and (b) $p_{\text{ctrl}}=0.364(4)$ with criticality consistent with random walk and percolation universality, respectively.
        The model is the local control with $x_F=\{0\}$ as shown in Fig.~\ref{fig:schematic}(f).
        The ensemble size is 2000, and the cut-off of zero is set to be $10^{-15}$.
    }
    \label{fig:linecuts_S0}
\end{figure}

Figure~\ref{fig:linecuts_S0} shows $I_3^{(0)}$ as a function of $p_{\rm ctrl}$ for various system sizes in the model of local control map with FM fixed point $x_F=\{0\}$.
In Fig.~\ref{fig:linecuts_S0}(a), we find that the crossing of $I_3^{(0)}$ occurs at a critical control rate of $p_{\text{ctrl}}=0.453(2)$, substantially larger than the value of $0.297(2)$ extracted from $I_3^{(1)}$ in Fig.~\ref{fig:linecuts}(e).
It is notable that the finite-size collapse saturates to a value $p_{\mathrm{ctrl}} = 0.453(2) < 0.5$, the expected value for control.
These transitions do coincide despite this numeric discrepancy, but in the full quantum numerics, we are limited by numerical precision (as was already seen in Ref.~\cite{zabalo2020critical}, see Appendix~\ref{app:S0_analysis} for a discussion).
Further evidence that these transitions coincide lies in the universal data: the correlation length critical exponent $\nu=1.10(6)$ is consistent with the random-walk universality of the CIPT.

Figure~\ref{fig:linecuts_S0}(b) plots the same quantities but with a finite projective measurement rate $p_{\text{proj}}=0.36$. 
We find that the critical control rate for $I_3^{(0)}$ now becomes $p_{\text{ctrl}}^c=0.364(4)$, which is separated from the CIPT at $p_{\text{ctrl}} \approx 0.5$.
Furthermore, $I_3^{(0)}$ is consistent with the percolation universality class with a critical exponent of $\nu\approx 1.25(9)$ ($\nu = 4/3$ for percolation). 

The zeroth Re\'nyi entropy exhibits a transition that coincides with the CIPT, both in its location and its universality, irrespective of the locality of the adder.
However, adding projective measurements still splits this entanglement transition from the CIPT.
These results are consistent with those obtained in Ref.~\cite{allocca2024statistical} for an effective statistical mechanics model in the limit of infinite onsite Hilbert space dimension, providing a helpful cross-check.

\subsection{Topology of the phase diagram}\label{sec:phase_diagram}

\begin{figure*}[ht]
    \centering
    \includegraphics[width=6.8in]{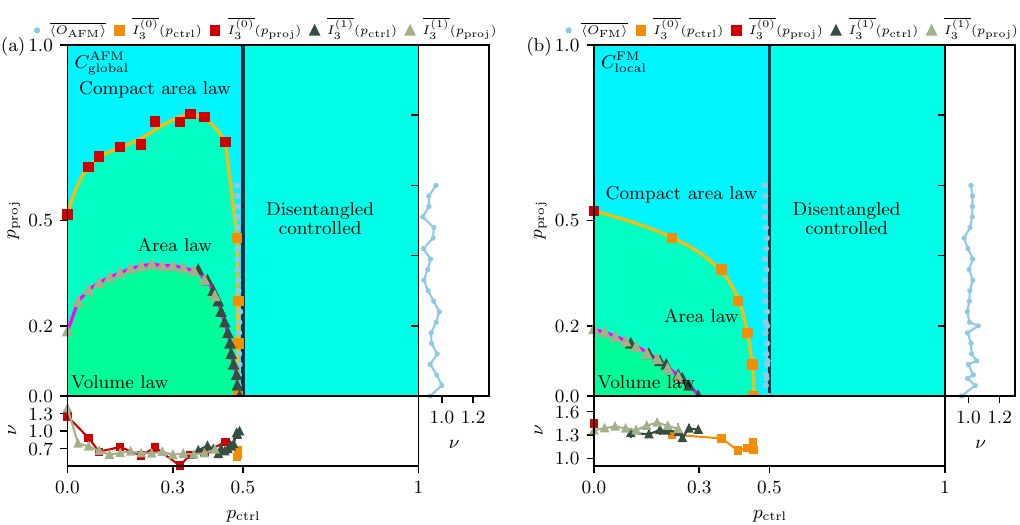}
    \caption{
    Phase diagram of the model for (a) the global adder with fixed points $x_F=\left\{ 1/3,2/3 \right\}$ [see Fig.~\ref{fig:schematic}(h)]; and (b) the local adder with a fixed point $x_F=\{0\}$ [see Fig.~\ref{fig:schematic}(i)].
    The markers indicate the critical points and exponents extracted from the data collapse, while the solid lines are for the schematic interpolation in the thermodynamic limit.
    The two types of tripartite mutual information $I_3^{(0)}$ and $I_3^{(1)}$ indicate their orthogonal directions in the data collapse. The ``compact area law'' refers to the phase where all R\'enyi entropies $n\ge0$ obey the area law.
    }
    \label{fig:pd}
\end{figure*}

The results of the previous sections motivate a thorough exploration of the full phase diagram as a function of the control rate and the rate of projective measurements.
In Fig.~\ref{fig:pd} we sweep both parameters, $p_{\text{ctrl}}$ and $p_{\text{proj}}$, to map out these phase diagrams for both the global adder with AFM fixed points $x_F=\left\{ 1/3,2/3 \right\}$ [Fig.~\ref{fig:schematic}(h)] and the local adder with FM fixed point $x_F=\left\{ 0 \right\}$ [Fig.~\ref{fig:schematic}(i)].
Each data point on each phase diagram boundary represents a critical point extracted by collapsing a particular data set (see Appendix~\ref{app:datacollapse} for details on our collapse methodology). 
The different colors represent different quantities used to detect the transitions, and whether the critical point was extracted from a vertical or horizontal sweep (i.e., of $p_{\rm proj}$ or $p_{\rm ctrl}$, respectively). 
The light blue dots correspond to the order parameters $\overline{\expval{O_{\text{AFM}}}}$ and $\overline{\expval{O_{\text{FM}}}}$. 
The light green and dark green triangular markers represent the tripartite mutual information $I_3^{(1)}$, extracted from vertical and horizontal sweeps, respectively. 
Similarly, the red and orange square markers indicate $I_3^{(0)}$, extracted from vertical and horizontal sweeps, respectively. 
The solid lines are schematic phase boundaries obtained from interpolating between the extracted critical points, with an extrapolation to the thermodynamic limit. 
The black solid line represents the CIPT, the magenta solid line represents the MIPT, and the orange solid line indicates the percolation transition witnessed by $I_3^{(0)}$.
We explain the two phase diagrams in more detail below.

\subsubsection{Global adder with $x_F=\left\{ 1/3,2/3 \right\}$}

The phase diagram for the global adder [Fig.~\ref{fig:pd}(a)] shows the coincidence of the CIPT (light blue dots) and MIPT (dark green triangles) at $p_{\text{proj}}=0$ to within our numerical accuracy. 
As $p_{\text{proj}}$ increases, the CIPT is not affected by the projective measurement, showing the same critical exponent $\nu\approx 1$ extracted from the order parameter $\overline{\expval{O_{\text{AFM}}}}$ (cyan dots, right subpanel).

However, the MIPT is gradually split from the CIPT, flowing to a lower critical control rate $p_{\text{ctrl}}^c$.
For larger projection rates $p_{\text{proj}}\gtrapprox 0.18$, the MIPT critical line assumes a dome shape, such that the area-law phase shows a re-entrance at small $p_{\text{ctrl}}$ [see also Fig.~\ref{fig:linecuts}(f)]. 
This unusual feature of an area-to-volume-law transition as $p_{\text{ctrl}}$ increases from 0 reflects the fact that the global adder can also contribute to generating entanglement entropy.
This unveils a nontrivial influence of the global adder. 
Namely, although the adder arises purely from a classical operation in the sense that it does not generate a quantum superposition when acting on a classical product state, it can generate entanglement entropy when applied to a quantum superposition.
We also find that the critical exponent of the MIPT changes from the Haar log-CFT universality class with $\nu \approx 1.3$ at $p_{\text{ctrl}}=0$, to the random walk universality class with $\nu\approx 1$ as it approaches the CIPT near $p_{\text{proj}}=0$ (see bottom subpanel).
Interestingly, along the bulk of the MIPT critical line separating volume- and area-law phases, the critical exponent $\nu$ exhibits a novel universality class of approximately $\nu \approx 0.7$.

This new universality class is intriguing and can also be observed in the percolation transition indicated by $I_3^{(0)}$, (red and orange data squares corresponding to vertical and horizontal line cuts, respectively).
We find that the percolation transition coincides with CIPT (the orange squares representing $I_3^{(0)}$ are invisible as they essentially overlap with the light blue dots representing $\overline{\expval{O_{\text{AFM}}}}$) in a large range of $p_{\text{proj}} \in [0,0.5]$.
Beyond $p_{\text{proj}}\gtrapprox0.5$, the percolation transition (red squares) starts to split from the CIPT and also manifests a dome shape, consistent with the MIPT critical line.
The critical exponent also changes from the percolation log-CFT universality class with $\nu\approx 1.3$ at $p_{\text{ctrl}}=0$ to this novel universality class with $\nu\approx 0.7$ along the percolation critical line (see bottom subpanel). 
However, one difference lies in the fact that $I_3^{(0)}$ (orange squares) does not manifest the random walk universality class with $\nu=1$ even when overlapping with the CIPT.

Having now identified this new universality class in the phase diagram, we can now search for its signature in the absence of additional projective measurements. Therefore,
using this as an initial guess we can perform a separate data collapse near $(p_{\text{ctrl}},p_{\text{proj}})=(0.5,0)$ that yields a different critical control rate of $p_{\text{ctrl}}=0.480(1)$ and critical exponent of $\nu=0.76(3)$ (see Appendix~\ref{app:dc_global}). 
These two critical control rates are too close to give an unbiased conclusion within the numerical accuracy of our simulations, leaving the possibility of a small area-law phase lying between CIPT and MIPT critical points.
We think that this second critical point at zero measurement rate could be reminiscent of the universality class with $\nu\approx 0.7$ at a finite measurement rate, which could contaminate the zero measurement rate case given the small system sizes restricted by the Haar random unitary circuit.

\begin{figure*}[ht]
    \centering
    \includegraphics[width=6.8in]{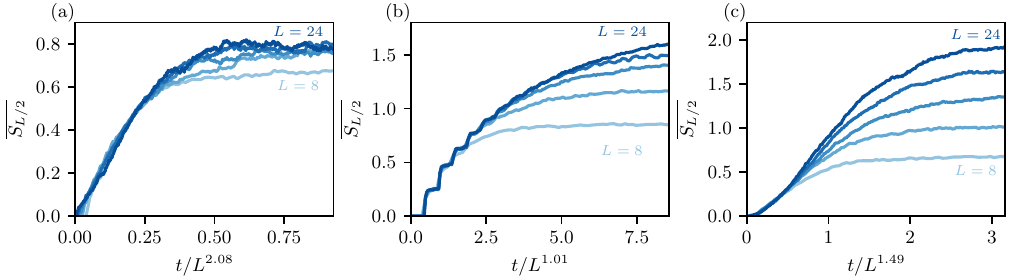}
    \caption{
        The half-cut entanglement entropy $\overline{S_{L/2}}$ as a function of time for a global control map with $x_F=\{1/3,2/3\}$ near (a) $(p_{\text{ctrl}},p_{\text{proj}})=(0.5,0.0)$ when $\nu\approx1$; (b) $(p_{\text{ctrl}},p_{\text{proj}})=(0,0.19)$ when $\nu\approx1.3$ (here the control rate is zero so it applies to regardless of the nature of control map); (c) $(p_{\text{ctrl}},p_{\text{proj}})=(0.42,0.3)$ when $\nu\approx0.7$. The dynamical exponents are (a) $z=2.08(3)$; (b) $z=1.01(2)$;  $z\approx1.49(2)$.
    }
    \label{fig:z}
\end{figure*}

We believe that this new universality class with $\nu \approx 0.7$ results from the global adder's nontrivial contribution to entanglement dynamics, as manifested in both $I_3^{(0)}$ and $I_3^{(1)}$.
To further characterize this novel universality class, we study its dynamical exponent. 
In Fig.~\ref{fig:z}, we plot the trajectory-averaged half-cut entanglement entropy $\overline{S_{L/2}}$ as a function of time for the global control map with $x_F=\{1/3,2/3\}$.
We fix $(p_{\text{ctrl}},p_{\text{proj}})=(0.42,0.3)$, where $I^{(1)}_3$ displays $\nu\approx0.7$.  
In the early-time regime $t\ll L$, we find that the growth of the half-cut entanglement entropy collapses to a single curve as $S\sim f(t/L^{1.49})$, indicating the dynamical exponent $z\approx1.49(2)$.
This dynamical exponent is distinct from the CIPT with $z=2$~\cite{iadecola2023measurement} [see Fig.~\ref{fig:z}(a)], and from the Lorentz-invariant value $z=1$ characterizing the Haar MIPT~\cite{zabalo2020critical} [see Fig.~\ref{fig:z}(b)].
We will leave more detailed studies of this exotic universality class with $\nu\approx0.7$ and $z\approx 1.49$ for future work. 

\subsubsection{Local adder with $x_F=\left\{ 0\right\}$}

For the local adder [Fig.\ref{fig:pd}(b)], the MIPT and CIPT correspond to two separate critical points, even at $p_{\text{proj}}=0$. The CIPT (light blue dots) remains at $p_{\text{ctrl}}^c\approx0.5$ with critical exponent $\nu\approx1$ (see right subpanel). 
However, as the projective measurement rate $p_{\text{proj}}$ increases, the MIPT phase boundary flows to smaller values of $p_{\text{ctrl}}^c$ until it reaches zero control rate. 
The critical point at $p_{\rm ctrl}=0$, $p^c_{\rm proj}\approx0.18$, is consistent with that found for the global adder at zero control rate [Fig.\ref{fig:pd}(a)]. 
Unlike with the global adder, the MIPT critical line for the local adder consistently displays Haar log-CFT universality with $\nu \approx 1.3$ [see triangles in bottom subpanel in Fig.\ref{fig:pd}(b)] and $z=1$ [see Fig.~\ref{fig:z}(b)].

The percolation transition witnessed by $I_3^{(0)}$ overlaps with the CIPT at $p_{\rm proj}=0$ and also flows to smaller $p^c_{\rm ctrl}$ as $p_{\rm proj}$ is increased [see the squares in Fig.~\ref{fig:pd}(b)].
At $p_{\rm ctrl}=0$, the value $p^c_{\rm proj}\approx0.5$ is consistent with the percolation transition in the standard Haar MIPT~\cite{skinner2019measurementinduced}.
The critical exponent $\nu$ also changes from $\nu\approx1$ at $p_{\rm proj}=0$ to the percolation log-CFT universality class with $\nu\approx 1.3$ as $p_{\text{proj}}$ increases [see squares in bottom subpanel in Fig.\ref{fig:pd}(b)]. This phase boundary agrees with that found in Ref.~\cite{allocca2024statistical} with the effective statistical mechanics model in the infinite onsite Hilbert space dimension limit.

\section{Interplay of Local and Global Control}\label{sec:nature}

\subsection{Entanglement effect of the adder}

In this section, we aim to understand how the locality of the adder affects the topology of the phase diagram.
Although the adder originates from a classical operation that does not generate quantum superpositions, it can create entanglement when acting on a superposition state. 
This occurs because the adder permutes wave function components among CB states. 
When applied to a CB state, this simply produces another CB state and does not produce entanglement.
However, when applied to a superposition state, it permutes multiple nonzero amplitudes, which generically affects the entanglement structure. 
For example, an adder which merely advances each CB state by 1 (i.e., $A=\sum_x\ketbra{x\oplus 1}{x}$) turns a disentangled state $\frac{1}{\sqrt{2}}\left( \ket{00}+\ket{01} \right)$ into an entangled Bell state $\frac{1}{\sqrt{2}}\left( \ket{01}+\ket{10} \right)$. 
In the chaotic phase, because of the extensive application of Haar random gates in the Bernoulli circuit, the system will typically be in a superposition state just before the adder is applied.
Another key distinction is that a local adder only affects an $O(1)$ number of qubits and therefore can only affect entanglement properties in a bounded region.
For this reason, we expect all models incorporating a local adder to be equivalent as they pertain to MIPTs, i.e., they should exhibit the same critical control rate $p_{\text{ctrl}}^c$ and critical exponent $\nu$ in the thermodynamic limit.
In contrast, the global adder can generate entanglement between disjoint subregions of the chain even with a single application.
Thus, we expect the global adder to have nontrivial implications for the entanglement dynamics, as we saw in the previous section where it leads to a new universality class of MIPT when projective measurements are added.

To make this argument concrete, we consider in Sec.~\ref{sec:local_AFM} a local adder that controls onto the same AFM fixed points $x_F=\{1/3,2/3\}$ as the global adder.
We then move on to study the interplay of the local and global adders in Sec.~\ref{sec:interpolation}.

\subsection{Local adder with $x_F=\left\{ 1/3,2/3 \right\}$}\label{sec:local_AFM}
\begin{figure}[ht]
    \centering
    \includegraphics[width=3.4in]{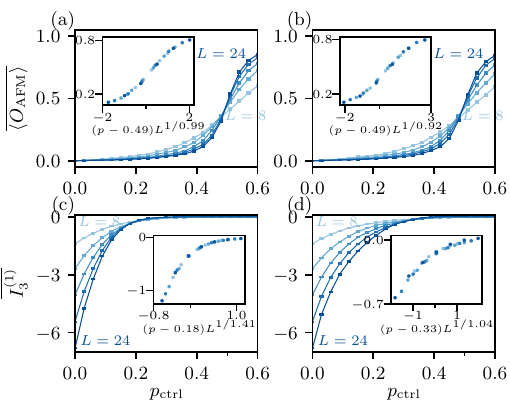}
    \caption{
    (a,c) The order parameter (Eq.~\eqref{eq:O_AFM}) and tripartite mutual information $\overline{I_3^{(1)}}$ [Eq.~\eqref{eq:I3}] for the model using the local adder with $x_F=\{1/3,2/3\}$ (i.e., $p_{\text{global}}=0$) [see Fig.~\ref{fig:schematic}(g)]. The insets show the data collapse near the critical point at (a) $p_{\text{ctrl}}=0.489(2)$ and (c) $p_{\text{ctrl}}=0.185(2)$ for $L\ge12$.
    (b,d) The interpolation between the local (Eq.~\eqref{eq:C_local_AFM}) and global adder [Eq.~\eqref{eq:C_global_AFM}] with $x_F=\{1/3,2/3\}$. The probability of the global adder is $p_{\text{global}}=0.15$ [see Fig.~\ref{fig:schematic}(j)].
    The insets show the data collapse near the critical point at (b) $p_{\text{ctrl}}=0.486(2)$ and (d) $p_{\text{ctrl}}=0.327(2)$ for $L\ge12$.
    The system sizes here range from $L=8$ to 24 and the ensemble size is 2000.
    }
    \label{fig:linecuts_int}
\end{figure}

We now consider the effect on the MIPT and CIPT of the local control map with $x_F=\{1/3,2/3\}$ defined in Sec.~\ref{sec:local_AFM} and illustrated in Fig.~\ref{fig:schematic}(g). 
Figures~\ref{fig:linecuts_int}(a) and (c) show the AFM order parameter and tripartite mutual information as functions of $p_{\text{ctrl}}$ for this new model featuring local control onto the AFM states.
Here, the two transitions split again, with the CIPT remaining at $p_{\text{ctrl}}^c\approx0.5$ and the MIPT happening at a lower value $p_{\text{ctrl}}^c=0.185(2)$, closer to its location in the bricklayer Haar model~\cite{zabalo2020critical}. 
We note that this value is substantially smaller than the critical control rate $p_{\text{ctrl}}^c=0.297(2)$ obtained in Figs.~\ref{fig:linecuts}(b,e) for local control onto the FM state.
This is because the local AFM control map involves an extra projective measurement on the first qubit [Fig.~\ref{fig:schematic}(d)].
This effectively increases the measurement rate by a factor of two, and indeed the MIPT occurs at about half the value of $p^c_{\rm ctrl}$ obtained for the local FM control.
The CIPT retains its original universality with $\nu=0.98(4)$, while the MIPT is again consistent with the Haar log-CFT with $\nu\approx 1.3$.

\subsection{Interpolation between the local and global adder}\label{sec:interpolation}
\begin{figure}[ht]
    \centering
    \includegraphics[width=3.4in]{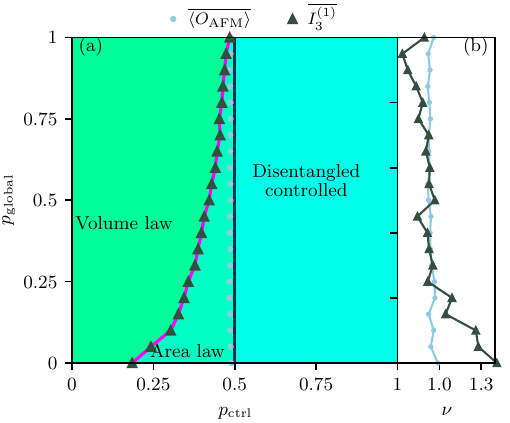}
    \caption{
    (a) The phase diagram of the model with the fixed points as $x_F=\{1/3,2/3\}$ interpolated between the local ($1-p_{\text{global}}$) and global adder ($p_{\text{global}}$).
    (b) The corresponding critical exponents $\nu$ for CIPT (cyan) and MIPT (magenta) as a function of $p_{\text{global}}$. 
    }
    \label{fig:pd_int}
\end{figure}
Since the global adder [Fig.~\ref{fig:linecuts}(a,d)] and the local adder [Fig.~\ref{fig:linecuts_int}(a,c)] have the same fixed points $x_F=\left\{ 1/3,2/3 \right\}$, we can randomly choose between the two adders to interpolate between them. 
We now modify the control protocol such that, at each control step, we apply the global control map with probability $p_{\text{global}}$ and the local control map with probability $1-p_{\text{global}}$.
Example order parameter and tripartite mutual information sweeps for $p_{\text{global}}=0.15$ are shown in Fig.~\ref{fig:linecuts_int}(b,d).
We find that, while the CIPT remains at $p_{\text{ctrl}}^c\approx0.5$ with the same critical exponent, the MIPT occurs at $p_{\text{ctrl}}^c=0.327(4)$, higher than the value $p_{\text{ctrl}}^c0.185(2)$ observed at $p_{\rm global}=0$ [Fig.~\ref{fig:linecuts_int}(a,c)].
Moreover, the MIPT acquires a critical exponent $\nu=1.046(7)$, in contrast to the value obtained for $p_{\text{global}}=0$ (that was consistent with Haar log-CFT universality). 
Thus, the two transitions remain split at finite $p_{\rm global}$, but they appear to manifest the same critical exponents. 
This suggests that the universality of the MIPT becomes more similar to that obtained for the global adder as $p_{\text{global}}$ increases.

To further characterize the interplay between the local and global adders, we sweep $p_{\text{global}}$ to map out the phase diagram as shown in Fig.~\ref{fig:pd_int}.
The critical line for the CIPT (black solid line) is obtained by collapsing the order parameter $\overline{\expval{O_{\text{AFM}}}}$, and the critical line for the MIPT (magenta solid line) is obtained by collapsing the tripartite mutual information $I_3^{(1)}$.
We find that the critical control rate $p_{\text{ctrl}}^c$ changes from 0.18 at $p_{\text{global}}=0$ to 0.5 at $p_{\text{global}}=1$, directly demonstrating that the global adder indeed drives the MIPT and CIPT together.
There is also a kink in $p_{\text{ctrl}}^c$ near $p_{\text{global}}=0.15$, beyond which the change in $p_{\text{ctrl}}^c$ becomes gradual.
In the right subpanel, we also present in Fig.~\ref{fig:pd_int}(b) the extracted critical exponents for both transitions as a function of $p_{\text{global}}$.
As $p_{\text{global}}$ increases, the MIPT critical exponent $\nu\approx1.3$ merges into that of the CIPT with $\nu\approx1$, which implies a change in the universality class from the Haar log-CFT to the random walk.
The most drastic change happens near $p_{\text{global}}=0.15$, which is also consistent with the kink in the $p_c$ as shown in Fig.~\ref{fig:pd_int}(a).
This trend of two critical exponents coalescing as the global adder is applied more frequently demonstrates that the universal features of the CIPT begin to overwhelm those of the MIPT as the two transitions draw closer to one another.

\section{Conclusion}\label{sec:conclusion}
In this work, we have performed a detailed study of control- and measurement-induced criticality in the quantum Bernoulli circuit, motivated by the general question of when the CIPT can herald the MIPT.
We find that the necessary condition for the MIPT and CIPT to coincide is to have a global control map that acts on the entire system, consistent with the results obtained in Ref.~\cite{sierant2023controlling} for Clifford circuits.
To do this, we first revisited the model of Ref.~\cite{iadecola2023measurement}, which features a global adder that controls onto the fixed points of $x_F=\left\{ 1/3,2/3 \right\}$. 
In this model, the CIPT and MIPT coincide at $p_{\text{ctrl}}\approx0.5$, and the MIPT inherits the universality class of the CIPT, which is described by a random walk with $\nu=1$ and $z=2$.
This demonstrates the possibility of using an observable in the CIPT to witness the MIPT.

We then showed that the CIPT and MIPT can be pulled apart by either replacing the global adder with a local adder or by introducing feedback-free projective measurements. 
In either case, the MIPT is pushed to a lower value of $p_{\rm ctrl}$, and we find that the MIPT critical exponent $\nu$ reverts to the Haar log-CFT universality class of $\nu\approx1.3$ and $z=1$; the CIPT remains fixed at $p^c_{\rm ctrl}=0.5$.
Adding the feedback-free projective measurements to the global-adder model allows us to smoothly track how the transitions split.
In this case, we find that the MIPT exhibits a novel universality class with $\nu\approx0.7$ and dynamical exponent $z\approx1.49(2)$ (with similar results in the zeroth Renyi entropy limit).
Thus, it appears that the interplay of nonlocal control and projective measurements strongly modifies the nature of the MIPT from the Haar log-CFT universality class observed in this work and elsewhere in the literature.

To demonstrate the effect of the locality of the adder on the topology of the phase diagram, we first proposed a new model with the same fixed points of $x_F=\left\{ 1/3,2/3 \right\}$ but with a local adder, and found that this change also makes the two transitions split. 
We then constructed a model that interpolates between the local and global adders by applying them randomly with a tunable bias. 
As the control is biased towards the global adder, the two criticalities come into closer proximity, and the critical properties of the MIPT become overwhelmed by those of the CIPT.

Several directions for future work present themselves.
It would be interesting to more thoroughly characterize the new MIPT universality class uncovered in this work that occurs in the presence of global control and projective measurements.
This could be done, e.g., using purification measures like those introduced in Ref.~\cite{gullans2020scalable}.
Moreover, it is worth investigating whether the difference in critical universality class relative to the standard Haar log-CFT can be traced back to properties of the area- and volume-law phases separated by the critical line, e.g., by studying subleading corrections to the entanglement entropy in the volume-law phase.
Finally, while this work has focused on the quantum aspects of the MIPT, the quantum nature of the CIPT has yet to be thoroughly investigated.
We expect quantum features of the CIPT to be most pronounced in the absence of projective measurements (see Fig.~\ref{fig:pd}), where the CIPT separates a dynamical phase with finite entanglement in the steady state from one that is completely disentangled. 

Experimentally, it will be exciting to try and witness each of these transitions separately and concomitantly by adjusting the nature of the feedback operations. Several experimental challenges remain, however, e.g., related to time delays of the reset operation in conjunction with mid-circuit measurements. Nonetheless, our work shows when the entanglement transition can be heralded by a linear-in-density-matrix observable and when non-linear quantities are necessary. 

\section*{Acknowledgement}
We acknowledge helpful discussions with Andrew Allocca, Kemal Aziz, Ahana Chakraborty, David Huse, Chao-Ming Jian, and Miles Stoudenmire.
This work is partially supported by Naval Research grant No.~N00014-23-1-2357 (H.P., J.H.P.), the National Science Foundation under Grants No. DMR-2143635 (T.I.),  No. DMR-2238895 (J.W.), and NSF CAREER Grant No. DMR-1944967 (S.G.).
The authors acknowledge the computing resources provided by the Office of Advanced Research Computing (OARC) at Rutgers, the Open Science Grid (OSG) Consortium.

\bibliography{Paper_CT.bib}
\appendix
\renewcommand{\thefigure}{\thesection\arabic{figure}}
\counterwithin{figure}{section}

\section{Data collapse}\label{app:datacollapse}
\subsection{Estimation of critical parameters}\label{sec:chi2}
We perform the data collapse of a set of $(p_i,y_i,L_i)$ to extract the critical point $p_c$ and critical exponent $\nu$ of the CIPT and MIPT. 
Here, $y_i=\frac{1}{N_{\alpha}}\sum_{\alpha}y_i^{(\alpha)}$ is the average of the actual metrics $y_i^{(\alpha)}$ (e.g., ${\expval{O_{\text{AFM/FM}}}}$ or ${I_3^{(n)}}$) for each quantum trajectory, $N_{\alpha}$ is the number of different trajectories $\alpha$.
$p_i$ is the control rate ($p_{\text{ctrl}}$) or projection rate ($p_{\text{proj}})$, and $L_i$ is the system size, which both are fixed given a specific data point. 

In the vicinity of the critical point, we expect all data points to fall on a single curve following the scaling form of $y_i=f((p_i-p_c){L_i}^{1/\nu})$, where $f$ is some unknown universal scaling function.

Therefore, we define $x_i=(p-p_c)L_i^{1/\nu}$ and sort $x_i$ in ascending order, to define the following loss function (reduced $\chi_\nu^2$ ) for the data collapse as
\begin{equation}\label{eq:chi2}
    \chi_{\nu}^2 = \frac{1}{N-2}\sum_{i=1}^{N}\left( \frac{y_i-y_i^\prime}{\sigma_i} \right)^2,
\end{equation}
Here, $N$ is the total number of data points with different $(p_i,L_i)$ (not to confuse with the total ensemble size $N_\alpha$), and $N-2$ is the total degrees of freedom given two fitting parameters $(p_c,\nu)$. $y_i^{\prime}$ is the linear interpolation between $y_{i-1}$  and $y_{i+1}$, i.e.,
\begin{equation}
    y_i^\prime = y_{i-1} + \frac{y_{i+1}-y_{i-1}}{x_{i+1}-x_{i-1}}(x_i-x_{i-1}).
\end{equation}
(For the two endpoints, we just use one-side interpolation.)
The estimated standard error of the mean $\sigma_i$ is 
\begin{equation}\label{eq:sigma}
    \sigma_i^2= \sigma_{{y_i}}^2 + \left( \frac{x_{i+1}-x_i}{x_{i+1}-x_{i-1}}\sigma_{{y_{i-1}}} \right)^2 + \left( \frac{x_{i-1}-x_i}{x_{i+1}-x_{i-1}}\sigma_{{y_{i+1}}} \right)^2
\end{equation}
where $\sigma_{{y_i}}^2$ is the standard error of the mean of $y_i$, and the rest terms are propagated from the error due to the interpolation between $y_{i-1}$  and $y_{i+1}$.

We minimize the loss function using the Levenberg-Marquardt algorithm~\cite{levenberg1944method,marquardt1963algorithm} to find the best fit of $p_c$ and $\nu$ with the Python package \texttt{lmfit}~\cite{newville2014lmfit} such that the reduced $\chi_{\nu}^2\approx1$.
If $\chi_{\nu}^2\gg 1$, it indicates underfitting, which is an indication of a bad choice of initial points of $(p_c,\nu)$ or the dataset itself has too much variance (probably due to the insufficient $N_\alpha$); however, if $\chi_{\nu}^2\ll 1$, it indicates overfitting of the variance of the data.

\subsection{Error estimation}\label{sec:Bootstrap}
The error bars of the critical point $p_c$ and critical exponent $\nu$ can be estimated directly from the square root of the diagonal element of the inverse of the Hessian matrix of the loss function at the optimum. 

However, one assumption using reduced $\chi_{\nu}^2$ is the normality of error (i.e., $\frac{y_i^{(\alpha)}-{y_i}}{\sigma_{y_{i}}}\sim\mathcal{N}(0,1)$), which may not be true for all metrics. 
In such case, one might need to normalize the data points of $\{y_{i}^\alpha\}$ through the Box-Cox transformation, however, it requires another level of choosing the parameter which could complicate the process of data collapse. Therefore, we use a simple while powerful technique that is insensitive to the underlying distribution of $\{y_{i}^\alpha\}$---bootstrapping.

For each point $(p_i, L_i)$, we will have a raw dataset of $\{y_{i}^\alpha\}$ corresponding to different quantum trajectories $\alpha$. 
We then resample $\{y_{i}^\alpha\}$ with replacement to obtain a new set of $\{y_{i}^{\tilde{\alpha}}\}$ for each $(p_i,L_i)$.
We perform the data collapse from Eq.~\eqref{eq:chi2} to Eq.~\eqref{eq:sigma} again to obtain a new estimate of $\tilde{p_c}$ and $\tilde{\nu}$.

We will repeat the process of resampling and refitting multiple times (practically, we find that 100 times suffice to converge) to obtain a set of $\{(\tilde{p_c},\tilde{\nu}),\dots\}$.
The error bar can be then estimated from the standard deviation of all the  $\{(\tilde{p_c},\tilde{\nu}),\dots\}$.
Finally, we combine the two error bars from both the inverse Hessian method in Sec.~\ref{sec:chi2} and bootstrapping, and choose the larger value as a conservative estimate for the error bar of the critical point and critical exponent.

\begin{figure}[ht]
    \centering
    \includegraphics[width=3.4in]{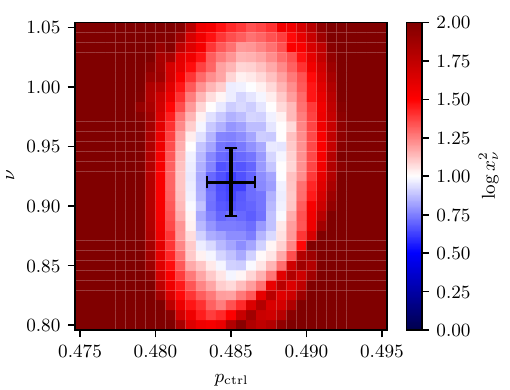}
    \caption{The profile of the reduced $\chi_{\nu}^2$ in the log scale as a function of $p_{\text{ctrl}}$ and $\nu$ for the data collapse in Fig.~\ref{fig:linecuts}(c). The range of the color bar only focuses on $[0,e^2]$, beyond which is saturated to $e^2$.
    Here, the error bar (black solid line) is estimated from the bootstrapping method.}
    \label{fig:chi2}
\end{figure}

We find that the bootstrapping method is more numerically stable than the inverse Hessian method alone. For example, in Fig.~\ref{fig:linecuts}(c), the inverse Hessian will give a very small error bar of $\sigma_{p_{\text{ctrl}}}=4\times10^{-4}$, and $\sigma_{\nu}=8\times10^{-3}$. 
However, from the profile of the reduced $\chi_{\nu}^2$ as a function of $p_{\text{ctrl}}$ and $\nu$ as shown in Fig.~\ref{fig:chi2}, we notice a broad region near the minimum. This ill-conditioned inverse Hessian can be remedied by the bootstrapping method, which gives an error bar of $\sigma_{p_{\text{ctrl}}}=2\times10^{-3}$, and $\sigma_{\nu}=3\times10^{-2}$.

\section{Kraus Operators description of all models}\label{app:kraus}
The Kraus operators for all six types of quantum circuits using the language of Kraus operators are shown in Table~\ref{table:kraus}.

\begin{table*}
    \caption{The Kraus operators for all six types of  quantum circuits in Figs.~\ref{fig:schematic}(e-j)}
    \label{table:kraus}
    \hspace*{-1.5cm}
    \begin{tabular}{lllllll}
    \hline
    Quantum circuit           & Fig.~\ref{fig:schematic}(e)    & Fig.~\ref{fig:schematic}(f)   & Fig.~\ref{fig:schematic}(g)   & Fig.~\ref{fig:schematic}(h)    & Fig.~\ref{fig:schematic}(i)   & Fig.~\ref{fig:schematic}(j)          \\ \hline
    Locality& Global & Local & Local & Global & Local & Interpolated \\ \hline
    Fixed points        & AFM    & FM    & AFM   & AFM    & FM    & AFM          \\ \hline
    Projection       & $\cross$     & $\cross$    & $\cross$   & $\checkmark$    & $\checkmark$    & $\cross$          \\ \hline
    Kraus Operators     &   {\tiny$\begin{matrix}
        \sqrt{1-p_{\text{ctrl}}} UT\\
        \sqrt{p_{\text{ctrl}}} A T^{-1}P_L^0\\
        \sqrt{p_{\text{ctrl}}} A T^{-1}P_L^1
    \end{matrix}$}       
    &   {\tiny$\begin{matrix}
        \sqrt{1-p_{\text{ctrl}}} UT\\
        \sqrt{p_{\text{ctrl}}} T^{-1}P_L^0\\
        \sqrt{p_{\text{ctrl}}} T^{-1}P_L^1
    \end{matrix}$}     
    &    {\tiny$\begin{matrix}
        \sqrt{1-p_{\text{ctrl}}} UT\\
        \sqrt{p_{\text{ctrl}}} T^{-1}X_LP_1^0 P_L^0\\
        \sqrt{p_{\text{ctrl}}} T^{-1} P_1^0 P_L^1\\
        \sqrt{p_{\text{ctrl}}} T^{-1} P_1^1 P_L^0\\
        \sqrt{p_{\text{ctrl}}} T^{-1}X_LP_1^1 P_L^1
    \end{matrix}$} 
       &    {\tiny$\begin{matrix}
        \sqrt{(1-p_{\text{ctrl}})(1-p_{\text{proj}})^2} UT\\
        \sqrt{(1-p_{\text{ctrl}})(1-p_{\text{proj}})p_{\text{proj}}} P_L^0 UT\\
        \sqrt{(1-p_{\text{ctrl}})(1-p_{\text{proj}})p_{\text{proj}}} P_L^1 UT\\
        \sqrt{(1-p_{\text{ctrl}})(1-p_{\text{proj}})p_{\text{proj}}} P_{L-1}^0 UT\\
        \sqrt{(1-p_{\text{ctrl}})(1-p_{\text{proj}})p_{\text{proj}}} P_{L-1}^1 UT\\
        \sqrt{(1-p_{\text{ctrl}})p_{\text{proj}}^2} P_{L-1}^0 P_L^0 UT\\
        \sqrt{(1-p_{\text{ctrl}})p_{\text{proj}}^2} P_{L-1}^0 P_L^1 UT\\
        \sqrt{(1-p_{\text{ctrl}})p_{\text{proj}}^2} P_{L-1}^1 P_L^0 UT\\
        \sqrt{(1-p_{\text{ctrl}})p_{\text{proj}}^2} P_{L-1}^1 P_L^1 UT\\
        \sqrt{p_{\text{ctrl}}} A T^{-1}P_L^0\\
        \sqrt{p_{\text{ctrl}}} A T^{-1}P_L^1
    \end{matrix}$}
        &    {\tiny$\begin{matrix}
            \sqrt{(1-p_{\text{ctrl}})(1-p_{\text{proj}})^2} UT\\
            \sqrt{(1-p_{\text{ctrl}})(1-p_{\text{proj}})p_{\text{proj}}} P_L^0 UT\\
            \sqrt{(1-p_{\text{ctrl}})(1-p_{\text{proj}})p_{\text{proj}}} P_L^1 UT\\
            \sqrt{(1-p_{\text{ctrl}})(1-p_{\text{proj}})p_{\text{proj}}} P_{L-1}^0 UT\\
            \sqrt{(1-p_{\text{ctrl}})(1-p_{\text{proj}})p_{\text{proj}}} P_{L-1}^1 UT\\
            \sqrt{(1-p_{\text{ctrl}})p_{\text{proj}}^2} P_{L-1}^0 P_L^0 UT\\
            \sqrt{(1-p_{\text{ctrl}})p_{\text{proj}}^2} P_{L-1}^0 P_L^1 UT\\
            \sqrt{(1-p_{\text{ctrl}})p_{\text{proj}}^2} P_{L-1}^1 P_L^0 UT\\
            \sqrt{(1-p_{\text{ctrl}})p_{\text{proj}}^2} P_{L-1}^1 P_L^1 UT\\
            \sqrt{p_{\text{ctrl}}} T^{-1}P_L^0\\
            \sqrt{p_{\text{ctrl}}} T^{-1}P_L^1
        \end{matrix}$}
           &       {\tiny$\begin{matrix}
            \sqrt{1-p_{\text{ctrl}}} UT\\
            \sqrt{p_{\text{global}}p_{\text{ctrl}}} A T^{-1}P_L^0\\
            \sqrt{p_{\text{global}}p_{\text{ctrl}}} A T^{-1}P_L^1\\
            \sqrt{(1-p_{\text{global}})p_{\text{ctrl}}} T^{-1}X_LP_1^0 P_L^0\\
            \sqrt{(1-p_{\text{global}})p_{\text{ctrl}}} T^{-1} P_1^0 P_L^1\\
            \sqrt{(1-p_{\text{global}})p_{\text{ctrl}}} T^{-1} P_1^1 P_L^0\\
            \sqrt{(1-p_{\text{global}})p_{\text{ctrl}}} T^{-1}X_LP_1^1 P_L^1
        \end{matrix}$}       \\ \hline
    \end{tabular}
\end{table*}

\section{Data collapse of the global control map at zero projection rate with another universality}\label{app:dc_global}

\begin{figure}[ht]
    \centering
    \includegraphics[width=3.4in]{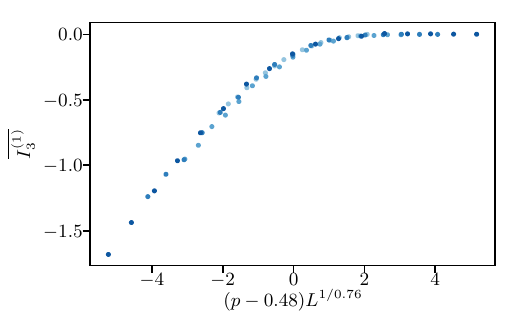}
    \caption{
    The data collapse of the tripartite mutual information for the global control map with $x_F=\{1/3,2/3\}$ at zero projection rate.
    The fitted critical control rate $p_{\text{ctrl}}=0.482(1)$ and $\nu=0.76(3)$ are different from the one in Fig.~\ref{fig:linecuts}(d).
    The reduced $\chi_{\nu}^2$ in Eq.~\eqref{eq:chi2} is 1.86, which is not statistically different from the $\chi_{\nu}^2\approx 1.83$ in the data collapse as shown in Fig.~\ref{fig:linecuts}(d).
    The system sizes here range from $L=8$ to 24 and the ensemble size is 2000.
    }
    \label{fig:linecuts_global}
\end{figure}

In Fig.~\ref{fig:linecuts_global}, we show that numerically the data collapses well for the global control map at zero projection rate with a modified universality class of $\nu\approx 0.7$, which is close to the universality class found in the finite projection rate case.

\section{Data collapse for the zeroth R\'enyi entropy}\label{app:S0_analysis}

In this section, we show a more systematic analysis of the data collapse for the zeroth R\'enyi entropy for the local adder at zero projection rate, corresponding to Fig.~\ref{fig:linecuts_S0}(a).

\begin{figure*}[ht]
    \centering
    \includegraphics[width=6.8in]{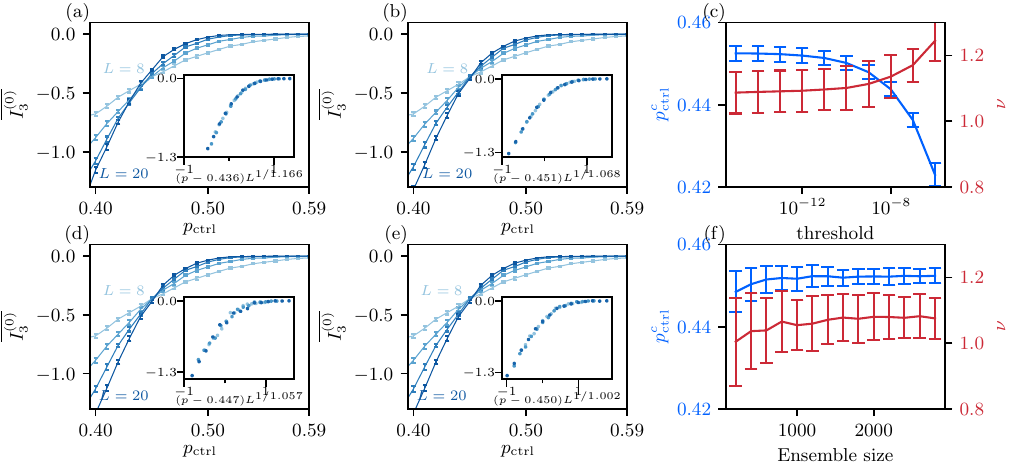}
    \caption{
        Top row: The data collapse of the tripartite mutual information using zeroth R\'enyi entropy for the local adder at zero projection rate (same as Fig.~\ref{fig:linecuts_S0}(a)) with (1) the threshold of $10^{-7}$ and (b) $10^{-11}$.  
        (c) The fitted critical control rate (blue, left axis) and critical exponent (red, right axis) as a function of the threshold.
        Bottom row: Same type of data, with a small ensemble size of (d) 500, and (e) 1000, using bootstrapping in Sec.~\ref{sec:Bootstrap}. 
        (d) The fitted critical control rate (blue, left axis) and critical exponent (red, right axis) as a function of the ensemble size.
    }
    \label{fig:linecuts_S0_extrapolate}
\end{figure*}

Since the zeroth R\'enyi entropy is very sensitive to the numerical threshold of zero (as it is the sum of all nonzero singular values in the Schmidt decomposition), in Fig.~\ref{fig:linecuts_S0_extrapolate}(a-b), we decrease the threshold from $10^{-7}$ to $10^{-11}$ to show that the drift of the critical control rate towards a higher value. 
The critical exponent also becomes closer to $\nu\approx 1$ of the random walk universality as we use a more stringent threshold.
Curiously, the critical exponent for lower thresholds is closer to the percolation value: hinting that our threshold is discarding truly nonzero eigenvalues of the reduced density matrix.
In Fig.~\ref{fig:linecuts_S0_extrapolate}(c), we present the two fitting parameters, $p_{\text{ctrl}}^c$ and $\nu$, as a function of the threshold, which manifests a clear trend of critical control rate increasing and critical exponent decreasing as the threshold decreases.
Despite the trend, there is not an asymptote to the control value of $p_{\mathrm{ctrl}} = 0.5$; this is consistent with earlier work where similarly-sized discrepancies were seen in the Hartley entropy as computed with thresholds in the full quantum dynamics \cite{zabalo2020critical}.
A less error-prone method would be to exactly keep track of bond dimensions such as within a matrix-product state.
Indeed, we speculate that such methods could be crucial for distinguishing the fate of $I_{3}^{(0)}$ in the non-local adder case.

In Fig.~\ref{fig:linecuts_S0_extrapolate}(d-e), we show the dependence of the ensemble size, which corresponds to the Monte Carlo sampling error.
Here, with a bootstrapping resampling, we increase the ensemble size from 500 to 1000, and find that the critical control rate and exponent do not change significantly.
However, their error bars decrease as the ensemble size increases, as shown in Fig.~\ref{fig:linecuts_S0_extrapolate}(f).

\end{document}